\documentclass[fleqn,usenatbib]{mnras}
\usepackage{graphicx,times}
\usepackage{natbib}
\usepackage{amssymb,amsmath}
\usepackage{hyperref}
\usepackage{soul}
\usepackage{epsfig}
\usepackage{multirow}
\usepackage{enumerate}
\usepackage{amsmath}
\usepackage{soul}
\usepackage{float}
\usepackage[section]{placeins}
\usepackage{url}
\usepackage{color}
\bibpunct{(}{)}{;}{a}{}{,}
\newcommand{\Rmnum}[1]{\uppercase\expandafter{\romannumeral #1}}

\usepackage{newtxtext,newtxmath}
\DeclareRobustCommand{\VAN}[3]{#2}
\let\VANthebibliography\thebibliography
\def\thebibliography{\DeclareRobustCommand{\VAN}[3]{##3}\VANthebibliography}
\usepackage{graphicx}	
\usepackage{amsmath}	
\usepackage{amssymb}	

\title[Two Origins of Radio Pulsars]
{Statistical tests of young radio pulsars with/without supernova remnants: implying two origins of neutron stars}

\author[X.H. Cui et al.]
{Xiang-Han Cui$^{1,2}$, Cheng-Min Zhang$^{1,2,3}\thanks{zhangcm@bao.ac.cn(CMZ)}$,
Di Li$^{1,2,4}$, Jian-Wei Zhang$^{1}$, Bo Peng$^{1,2,5}$, Wei-Wei Zhu$^{1,2}$,
\newauthor  Qing-Dong Wu$^{6}$,  Shuang-Qiang Wang$^{6}$, Na Wang$^{6}$, De-Hua Wang$^{7}$, Yi-Yan Yang$^{8}$, Zhen-Qi Diao$^{7}$,
\newauthor Chang-Qing Ye$^{7}$, and Hsiang-Kuang Chang$^{9}$\\
$^1$National Astronomical Observatories, Chinese Academy of Sciences, Beijing 100101, China\\
$^2$School of Astronomy and Space Science, University of Chinese Academy of Sciences, Beijing 100049, China\\
$^3$School of Physical Sciences, University of Chinese Academy of Sciences, Beijing 100049, China\\
$^4$NAOC-UKZN Computational Astrophysics Centre, University of KwaZulu-Natal, Durban 4000, South Africa\\
$^5$Guizhou Radio Astronomy Observatory, Chinese Academy of Sciences, Guiyang 550025, China\\
$^6$Xinjiang Astronomical Observatory, Chinese Academy of Sciences, Urumqi, Xinjiang 830011, China\\
$^{7}$School of Physics and Electronic Science, Guizhou Normal University, Guiyang 550001, China\\
$^{8}$School of Physics and Electronic Science, Guizhou Education University, Guiyang 550018, China\\
$^{9}$Institute of Astronomy, National Tsing Hua University, Hsinchu 30013, Taiwan, China
}
\date{Accepted XXX. Received YYY; in original form 2021}

\pubyear{2021}

\begin{document}
\label{firstpage}
\pagerange{\pageref{firstpage}--\pageref{lastpage}}
\maketitle

\begin{abstract}
The properties of the young pulsars and their relations to the supernova remnants (SNRs) have been the interesting topics.
At present, 383 SNRs in the Milky Way galaxy have been published,  which are associated with
64 radio pulsars and 46 pulsars with high energy emissions.
However, we noticed that 630 young radio pulsars with spin periods of less than half a second
have been not yet observed the SNRs surrounding or nearby them,
which arises a question of that could the two types of young radio pulsars with/without SNRs hold distinctive characteristics?
Here, we employ the statistical tests on the two groups of young radio pulsars with (52) and  without (630) SNRs to reveal if they share different origins.
Kolmogorov-Smirnov (K-S) and Mann-Whitney-Wilcoxon (M-W-W) tests indicate that the two samples have the different distributions
with parameters of spin period ($P$), derivative of spin period ($\dot P$), surface magnetic field strength ($B$), and energy loss rate ($\dot E$).
Meanwhile, the cumulative number ratio between the pulsars with and without SNRs at the different spindown ages decreases significantly after $\rm10-20\,Kyr$.
So we propose that the existence of the two types of supernovae (SNe), corresponding to their SNR lifetimes, which can be roughly ascribed to the low-energy and high-energy SNe.
Furthermore, the low-energy SNe may be formed from the $\rm8-12\,M_{\odot}$ progenitor,
e.g., possibly experiencing the electron capture, while the main sequence stars of $\rm12-25\,M_{\odot}$ may produce the high-energy SNe probably by the iron core collapse.

\end{abstract}

\section{Introduction}\label{1}
The associations between the pulsars and supernova remnants (SNRs)  have been
of considerable interest topics in neutron stars (NSs) astrophysics, since
the radio pulses were firstly observed in the Crab Nebula in 1968 \citep{Staelin68},
as well as  the discovery of the Vela pulsar \citep{Large68}.
The story is continuing  with identifying the  potential NS in SN 1987A \citep{Page20, Greco21, Soker21}.
Thanks the efforts and  developments
of astronomical  facilities in recent years, the number of pulsars and SNRs has increased significantly.
Up to now, there are more than 3000 pulsars (ATNF\footnote{https://www.atnf.csiro.au/research/pulsar/psrcat/}: 2781, GPPS\footnote{http://zmtt.bao.ac.cn/GPPS/}: 201 \citep{Han21}, CRAFT\footnote{https://crafts.bao.ac.cn/}: 125) observed in the
radio band and 383 SNRs (SNRcat\footnote{http://snrcat.physics.umanitoba.ca/SNRtable.php}) in the Milky Way galaxy had been published \citep{Ferrand12}.
Among these, there are 110 pulsars that have been identified in SNRs, including 6 anomalous X-ray pulsars (AXPs), 5 soft gamma-ray repeaters (SGRs), a total of 13 magnetar candidates, and 15 central compact objects (CCOs) or CCO candidates.
By comparing with ANTF Pulsar Catalogue, there are also 18 pulsars without radio emissions (NRAD) in SNRcat, only observed at the infrared or higher frequencies \citep{Manchester05}.
In short, 64 radio pulsars with SNRs have been published, as seen in Table 1.
Interestingly, the number  ratio between the radio pulsars (64) and   SNRs (337)  is about 1/5, which is consistent with the estimation by the beaming fraction of radio pulsars \citep{Taylor77, Lorimer93, Lorimer12}.
\begin{table}
\centering \caption{List of various types of pulsars with SNRs}
\begin{tabular}{@{}lcc@{}}
\hline
\noalign{\smallskip}
\bf Source  &\bf Number       &\bf Ref. \\
\hline
\noalign{\smallskip}
SNR & 383 & [1] \\
Pulsar$^a$ & 110 & [1] \\
Radio pulsar & 64$^b$ & [2] \\
Magnetar or candidate  & 13$^c$ & [1, 3, 4] \\
CCO$^d$ or candidate  & 15 & [1] \\
NRAD$^e$ & 18 & [2] \\
\hline
\end{tabular}
\label{tab1}
\begin{flushleft}
{\bf Note:\\}
$^a$ Various pulsars with SNRs.

$^b$ The total number of radio pulsars with SNRs is 64, but
only 52 of them  are analyzed in this article (selection details
in Section 2.1).

$^c$ AXP (anomalous X-ray pulsar): 6, SGR (soft gamma-ray repeater): 5.
{\url{http://www.physics.mcgill.ca/~pulsar/magnetar/main.html}}

$^d$ CCO: central compact object.

$^e$ Pulsars without radio emissions and do not contain the
magnetars, CCOs, and their candidates.

{\bf Ref.}: [1] \cite{Ferrand12}; [2] \cite{Manchester05}; [3] \cite{Olausen14}; [4] \cite{Esposito21}.
\end{flushleft}
\end{table}

From Table 1,  the question of why only a fraction of SNRs have been found with pulsars  can be solved by their  beaming cone angles across the
earth.
Meanwhile,   some other explanations are worthy of mentioning as pointed out as below.
First of all, not every SNe could generate a NS  or some NSs might  not be detectable as pulsars \citep{Radhakrishnan80, Srinivasan84, Manchester87, Narayan88}.
Next, the flux density or luminosity of young radio pulsars may be overestimated, implying  that some young radio
 pulsars are too faint to be observed in some SNRs \citep{Stollman87, Lorimer93}.
Finally, the  kick velocity of pulsar may be quite high after   birth, which could  result in the pulsars to escape from their  SNRs \citep{Frail94}.

Although some pulsars can be found in the pulsar wind nebulae \citep{Gaensler06},  it is still an open question that  so
many young radio pulsars have not seen their SNRs.
Nowadays, there exist 630 young radio pulsars (spin period less
than 0.5\,s, and details as described in Section 2) without SNRs,
which provides  us a new aspect to further study the association
between  pulsars and SNRs, as well as the NS origins. In the
former studies of X-ray pulsars \citep{Knigge11} and double
neutron star (DNS) \citep{Yang19} population, researchers
suggested  that the NSs may birth from the different  origins,
i.e., the existence of the electron capture and iron core collapse
SNe \citep{Janka12}. In the recent work about pulsars and SNRs,
\cite{Malov21} noticed that the mean values  of radio luminosity
of pulsars observed inside and outside SNRs are significantly different with one order of magnitude.

Inspired by these researches, we conjecture that there may exist
two origins for radio pulsars. Therefore, we attempt to use the
statistical methods to study the properties of  radio pulsars  and
their relations with SNRs. We apply some physical criteria to
select two samples of pulsars, that is, the radio pulsars with SNR
(SNR-PSRs, 52) and without SNR (non SNR-PSRs, 630) (see
Section 2.1 for details) By drawing the cumulative distribution
function (CDF) and utilizing the Kolmogorov-Smirnov (K-S) and
Mann-Whitney-Wilcoxon (M-W-W) tests for these two samples, we find
that the obtained  results support the  different distributions
for  two samples. 
Based on spin period evolution model, we estimate the spindown age (not characteristic age) of radio pulsars in these two samples.
After a further discussion, it is inferred that the two samples of pulsars may origin from two types of progenitors, such as the low-energy and high-energy SNe (e.g., electron capture and iron core collapse), respectively.
However, the energy boundary is still unclear, or there may be an overlapping part between them,
because some researches in SNRs indicated that SNRs may have a continuous energy distribution like lognormal \citep{Leahy17, Leahy20}.
Meanwhile, we notice that the cumulative number ratios  of SNR-PSRs to  non SNR-PSRs are
decreased quickly after $\rm10-20\,Kyr$. Finally, according to the initial mass function (IMF) or Salpeter function
\citep{Salpeter55}, it is possible to statistically distinguish the two types of their progenitor stars at the mass boundary of
$\rm\sim 12\,M_\odot$.

The structure of our paper is presented as follows.
In Section 2, we describe the data selection of pulsars and introduce the spin period evolution model.
In Section 3, we apply the statistical tests on the two samples  and analyze the results.
Finally, in Section 4, we discuss a possible physical significance for the two distributions of the pulsars,
and main conclusions are summarized also.

\section{Data and Model}\label{2}
\subsection{Data Selection}
Our data are taken from ATNF Pulsar Catalogue \citep{Manchester05} and SNRcat \citep{Ferrand12}.
Data selections are made to construct the two samples of the  young radio pulsars with and  without SNRs, and
the  selection criteria are described below.

1) We choose the pulsar samples in the  two data bases with the spin periods ($P$) less than 0.5\,s.
Due to the short duration of SNRs, the pulsars with SNRs are taken as the young pulsars \citep{Gaensler95, Malov21}.
Considering that the characteristic age of pulsar usually has a significant error compared
 to the real age \citep{Lai96, Kaspi01, Tian06, Lyne12}, we apply the spin period to represent the age under the  spin period evolution model (details in Section 2.2).
Generally speaking, the lifetime of SNRs usually less than 300\,Kyr, so the  spin period of pulsars may be around 0.5\,s based on the magnetic
dipole model (see Section 2.2 and Appendix A for calculation details).
Meanwhile, for all radio pulsars, the median of period is about 0.5\,s, which can be known  from ATNF database \citep{Manchester05}.
Additionally, because the periods for most rotating radio transients (RRATs) and intermittent pulsars are greater than 0.5\,s \citep{McLaughlin06, Kramer06}, we   eliminate these special radio pulsars from our samples.
Therefore, we regard the radio pulsars  with   spin periods of  less than 0.5\,s  to be the young pulsar, and they are roughly consistent with the
arguments based on that their magnetic fields are comparable  with those measured by cyclotron absorption lines of X-rays \citep{Ye19}.

2) The data with the surface magnetic field strength ($B$) ranged  from $\rm 10^{11} \, G$ to $\rm10^{14} \, G$ are used. If the pulsar's B-field is lower than $\rm10^{11}\, G$, it may be a millisecond pulsar (MSP) \citep{Bhattacharya91, Lorimer08} or a CCO \citep{Halpern10, Gotthelf13}. If the pulsar's B-field is higher than $\rm10^{14} \, G$, it may be a magnetar \citep{Duncan92, Ferrario08, Kaspi17, Esposito21}.

3) We select the data to satisfy that their spin periods and magnetic fields are distributed  above the  spin-up line in B-P diagram  \citep{Bhattacharya91}. While, the pulsars below the spin-up line may have a more complicated evolutionary tracks, such as experiencing
the accretion spin-up in binary systems \citep{Zhang06}.

4) Some types of the special pulsars are removed from our samples,  including the MSPs, magnetars, recycled pulsars \citep{Zhang06}, CCOs, RRATs, intermittent pulsars and NRADs.  The physical characteristics between the young radio pulsars and these  special pulsars are significantly different, so they may follow the different  birth conditions and evolutionary paths.

After setting these selections ($ P<0.5s$, $10^{11}G<B<10^{14}G$, above the spin-up line, and removing the
special pulsars), we obtain the two groups of  samples of 52
SNR-PSRs and 630 non SNR-PSRs, respectively, as illustrated in
Figure 1.
\begin{figure}
\centering
\includegraphics[width=7.8cm]{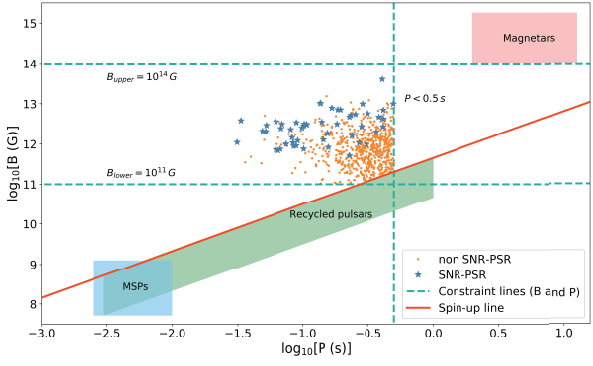}
\caption{Data distribution graph in the magnetic field versus spin period (B-P) diagram after the selection criteria. The blue stars and orange dots stand for the SNR-PSRs and non SNR-PSRs, respectively.
The green dashed lines represent the  constraint lines of surface B-field strength (B) upper ($B_{upper}=\rm10^{14}\,G$) and lower ($B_{lower}= \rm10^{11}\,G$) limit and spin period ($P<0.5\,\rm s$).
The red solid line represents the spin-up line of accretion   pulsars in binaries (Bhattacharya and van den Heuvel 1991).
The red, green and blue area stand for  the approximate ranges of the magnetars, recycled pulsars   and millisecond pulsars (MSPs).}
\label{fig1}
\end{figure}

\subsection{The Model}
Because of  employing  the spin period to represent the pulsar's age (spindown age),
we  briefly introduce the model\citep{Shapiro83, Camilo94, Lorimer12, Lyne12}.
The energy loss rate of a spin-powered pulsar can be expressed as, $\dot E = {dE}/{dt}={d(I\Omega^2/2)}/{dt}=-I\Omega\dot\Omega$, where $E$ is total energy of pulsar, $\dot E$ is energy loss rate, $\Omega$ and $\dot \Omega$ are the spin angular velocity and its  derivative, and $I$ is moment of inertia \citep{Shapiro83}.
The pulsar kinetic energy and the radiation energy loss rate are equal, then we have, $I\Omega\dot\Omega=k\Omega^{\rm n+1}$, with $k$  a coefficient, where the braking index n=3 for  the magnetic dipole model.
The spin period evolution equation can be written as (calculation details in Appendix A)
\begin{equation}
\begin{split}
P(t)=P_0\left( \frac t \tau_0 + 1\right)^{1/(\rm n-1)},
\end{split}
\end{equation}
where $P_0$,   $\dot P_0$, and $\tau_0$  are the  spin period, the derivative of spin period and the spindown age at present, respectively.
The spindown age of a pulsar is defined as  $\tau={P} /{[({\rm n-1})\dot P]}$ \citep{Camilo94}.
When $t\gg\tau_0$ ($P\gg P_0$), the spin period evolution equation can be expressed as
\begin{equation}
\begin{split}
P(t)\approx P_0\left( \frac t \tau_0 \right)^{1/(\rm n-1)}.
\end{split}
\end{equation}
With n=3 and parameter $k=-{B_p^2R^6}/{(6c^3)}$ is obtained by the
magnetic dipole model, where $B_p$ is polar magnetic, $R$ is NS
radius, and $c$ is speed of light \citep{Shapiro83}. Then the
Eq.(2) can be written as $P(t)\approx P_0( t /\tau_0)^{1/2}$.
Because n=3 is a theoretical value and the measured values of many
pulsars are generally less than it \citep{Johnston99, Kou15}, we
take n=2.7 to estimate the ages of pulsars, and the reasons are
interpreted as follows. Firstly, \cite{Lyne85} and
\cite{Lorimer04} estimated that the number of observable pulsars
in Milky Way galaxy is $\sim 2-7 \times 10^4$, which corresponds
to  the relevant life time of radio pulsars to  be  $\rm\sim
20\,Myr$. If the braking index n is too small, like n=2.1,
the corresponding life time of pulsars will decrease to one or two
million years to reach the observational limit spin period of
about 10 s. This will result in too few observable pulsars, which
is inconsistent with the observational facts at present. Therefore, it
is the reason that we do not apply the Crab pulsar's breaking
index as an average value for the whole pulsar samples, which is
from 2.1 to 2.5 or 2.6 \citep{Lyne15, cadez16}. Secondly, almost
all radio pulsar periods  are less than 10s (only two are longer
than 10s, e.g., J2251-3711 with 12.1s \citep{Morello20} and
J0250+5854 with 23.5s \citep{Tan18}). By considering the above two
reasons, we take the braking index n=2.7 as a statistical
value to estimate the evolutionary time, which can give the spin
period to be about 10s after evolving about 20 Myrs. Then we
select  the initial conditions of the Crab-like pulsars to perform
the calculations, e.g.,  $P_0 =0.033$\,s and $\dot P_0 =
4.21\times 10^{-13}\,\rm ss^{-1}$ \citep{Goldreich69, Lyne15},
implying $\tau_{02.7}=1462$\,yr (n=2.7) and $\tau_{03}=1243$\,yr
(n=3, characteristic age). The spin period evolutionary
curves are plotted in Figure 2 based on Eq. (2). Here, we
point out that, although  initial spin periods of neutron stars
may be not like that of the Crab pulsar's \citep{Popov12,
Igoshev13}, we still insist to employ the Crab pulsar as a
reference sample since it is the sole  pulsar with the known real
age.
\begin{figure}
\centering
\includegraphics[width=7.8cm]{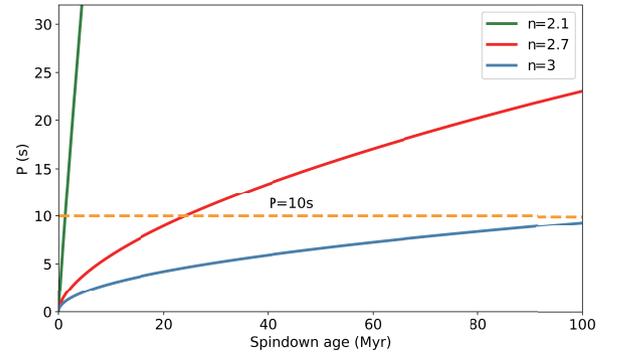}
\caption{The diagram of spin period evolution with various
braking index. Different colors label  the evolution curves with
the  breaking index n=2.1 (green), n=2.7 (red), and  n=3 (blue),
respectively. The orange dashed horizontal line shows the required
time for a pulsar period evolving  to 10s.} \label{fig2}
\end{figure}

\section{Statistics and Results}\label{3}

In the following analysis, we employ the two  statistical tests,   K-S and M-W-W  \citep{Yang19, Cui21}, to check
whether two samples hold  the same distribution.
The test results are represented by the parameters, p-values, and the procedure is described in the  following.
If p-value is less than 0.05, it indicates that this  test rejects the null hypothesis  (the two samples have the same distribution) at 5\% significance level.

\begin{figure}
\centering
\includegraphics[width=7.8cm]{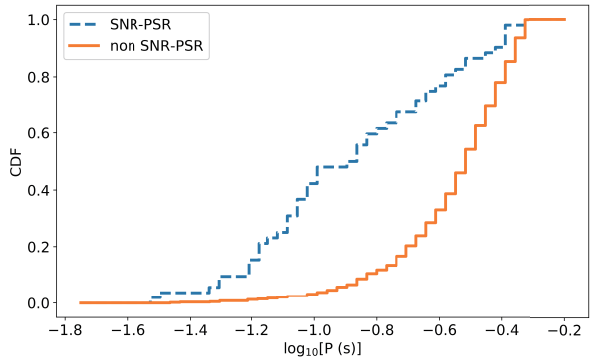}
\caption{The Cumulative distribution function (CDF) of spin period for  SNR-PSRs and non SNR-PSRs. The dashed (solid) line stands  for SNR-PSRs (non SNR-PSRs).}
\label{fig3}
\end{figure}

In order to see whether the two groups of samples of SNR-PSRs and non SNR-PSRs hold the same distribution, we  draw the  CDF curves
in Figure 3, where  the two curves are conspicuously separated to each other.
To show the difference of the two samples  more quantitatively, we apply the K-S test and M-W-W test, and
the p-values of these two tests are  as  low as $1.98\times10^{-12}$ and $5.85\times10^{-13}$.
The very low p-values indicate that the distributions of the two samples  should have   the different origins.
Therefore, with the results of two CDF curves and p-values, we believe that the two samples may come  from the
different statistical distributions. The further test results for the other parameters  are shown in Appendix B.

\begin{figure}
\centering
\includegraphics[width=7.8cm]{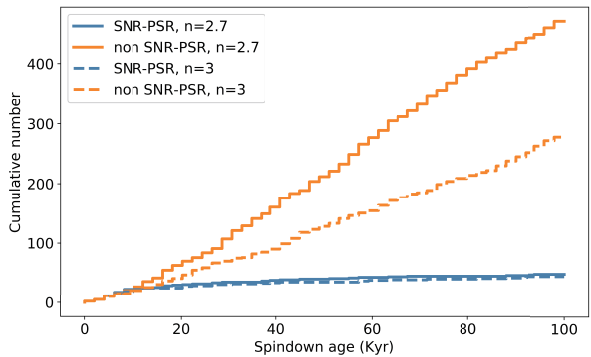}
\caption {The cumulative number distribution of SNR-PSRs and non SNR-PSRs in different ages.
The blue (orange) solid lines stand for SNR-PSRs (non SNR-PSRs) under n=2.7, and  dashed lines stand for the cases of n=3.}
\label{fig4}
\end{figure}

Next, we draw a cumulative number distribution (Figure 4) for the  different ages that are  estimated  by Eq.(2).
Interestingly, for the age less than  $\sim$ 10\,Kyr the two curves of SNR-PSRs and non SNR-PSRs  coincide together, however,
after $\rm10-20\,Kyr$, the two curves  are  drifted  away.
The cumulative number ratio,  expressed as $N_{snr}/N_{nsnr}$,
for the  two samples with  the different ages probably can infer a fact that the two samples hold the different origins.

In order to see the variation of this  ratio between two samples more intuitively, we plot the different ratio values from the age of
 5\,Kyr to 100\,Kyr in Figure 5. We find that,  before the age of  25\,Kyr, we obtain 5 ratio points for n=2.7 and n=3.
While,  from 25\,Kyr to 100\,Kyr, only 2 ratio points are obtained  because of the inadequate data.
From the data point of these ratio values, we obtain  that the number ratio between SNR-PSR and non SNR-PSR is  close to unity
at  the age of  $\sim$ 10\,Kyr.
However, after $\rm10-20\,Kyr$, there exists a sharp decline in the ratio values.
Meanwhile, for the ratio value  after $\sim$ 10\,Kyr,  with n=2.7 (n=3),  we obtain a relation between the ratio and the age  $\mathcal{\phi}_{2.7}=1.1(t/10kyr)^{-1.01}$ ($\mathcal{\phi}_3=1.2(t/10kyr)^{-0.91}$) and goodness of fit $\mathcal{R}_{2.7}^2=0.998$ ($\mathcal{R}_3^2=0.996$), as described in figure 5.

\begin{figure}
\centering
\includegraphics[width=7.8cm]{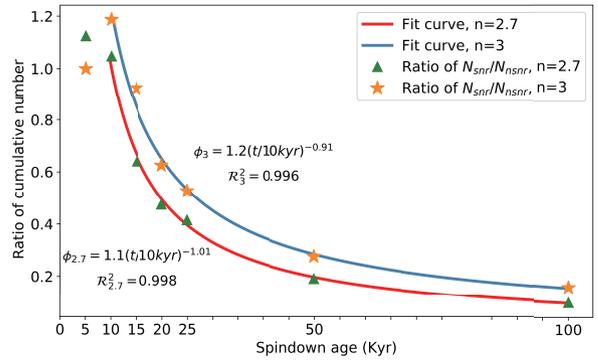}
\caption{The evolution diagram of cumulative number ratio between SNR-PSRs ($N_{snr}$) and non SNR-PSRs ($N_{nsnr}$). With n=2.7 (n=3), the orange stars (green triangles) stand for the ratio in different ages, and the solid blue (red) line is the fitting curve of these ratio points.}
\label{fig5}
\end{figure}

\section{Discussions and Conclusions}\label{4}
On the reasons for a sharp drop of the cumulative number ratio between the SNR-PSRs and non-SNR-PSRs, after $\rm10-20\,Kyr$, together with the
the different distributions,  we think that there may exist  two ways of birth  for radio pulsars.
The long aged SNR-PSRs may be involved in the high-energy SNe, whereas the non SNR-PSRs may be related to the low energy cases with short duration.
For high-mass stars, their SNRs may survive longer time.
Correspondingly, the low-energy SNe  may be involved in the explosions of the low-mass progenitor stars, in which the SNRs may last a shorter duration.

However, can the effect of decline in cumulative number ratio is caused by the  age difference?
For example, \cite{Leahy20} analysed a 15-Galactic-SNR sample with the best-fit mean energy of $\rm2.7\times 10^{50}\,erg$,
and they concluded that SNRs become incomplete and hard to identify after 30 Kyr.
In order to discuss this question more clearly, we need to test for samples with the  different spindown ages.
According to Figures 4 and 5, the two curves are separated after about $\rm10-20\,Kyr$ for various braking index.
So, we can take 10 Kyr as the critical  boundary and divide SNR-PSRs and non SNR-PSRs into two groups  by  the ages of  less and greater than 10 Kyr.
For the pulsar ages less than 10 Kyr, the $P$ distributions of  SNR-PSRs and non SNR-PSRs are same, while their $\dot P$ distributions are  different.
However, for the pulsar ages more than 10 Kyr, both distributions of $P$ and $\dot P$ are different (details in Appendix C).
These results may infer that their initial periods are independent on the types of pulsars, but their braking mechanisms are different. %
The birth of the NSs spin are due to the transfer of angular momentum of progenitors to NSs \citep{Lyne12}, then the energy of SN explosions might  little effect on their spin periods.
Meanwhile, the difference in $\dot P$ will affect the evolution of $P$, resulting in a different distribution of $P$ after 10 Kyr.
The above evidence support that their intrinsic differences may lead to different distribution among SNR-PSRs and non SNR-PSRs.
Specifically, when radio pulsars are just born, the differences of progenitor mass or explosion energy may create differences in the two groups.
The effect of age is to amplify the differences during the evolution, which origin from different initial $\dot P$.
Thus, this indicates that the differences between SNR-PSRs and non SNR-PSRs in Figures 3, 4, and 5 are more likely to be caused by multiple reasons, rather than only an age difference.
The reasons may include different neutron star generation mechanisms and evolution over time.

We need to emphasize that the duration of $\rm10-20\,Kyr$ is not a strict time, but a statistical value.
The result based on SNR evolutionary models \citep{Leahy20} may be slightly different from our result.
Although our results ($\rm10-20\,Kyr$) are not completely the same with their (30 Kyr),
at least it shows a life time boundary of two type SNRs, no matter from the perspectives of pulsars in radio and SNRs in X-ray.
It can be inferred from this boundary that two types of pulsars generated by two types of SNRs can be roughly distinguished.
Therefore, the cumulative number ratio of $\sim 1$ at $\sim$ 10\,Kyr may represent the ratio of these two kinds of pulsar production, that is
\begin{equation}
\begin{split}
\psi = \frac{N_{snr}}{N_{nsnr}}\sim\frac{N_{high-energy}}{N_{low-energy}}\sim\frac{N_{high-mass}}{N_{low-mass}}\sim 1,
\end{split}
\end{equation}
where $N_{snr}$ ($N_{nsnr}$), $N_{high-energy}$ ($N_{low-energy}$) and $N_{high-mass}$ ($N_{low-mass}$) represent the numbers
of SNR-PSRs (non SNR-PSRs), high (low)energy SNe, and  high (low) mass stars.
Specifically for the Crab pulsar, it may born from a low-energy SN ($\sim\rm 10^{50}\,erg$) and low-mass progenitor star \citep{Yang15}.
However, the earlier view believed that the Crab Nebula has a more extended remnant \citep{Chevalier77}, which indicates a higher energy ($\sim\rm 10^{51}\,erg$).
Although no medium have been detected in the surrounding area at radio or X-ray band \citep{Frail95, Seward06},
it also reminds us that total energy of the Crab Nebula is still an open question.

The mass range of the progenitor stars for NS formations approximately lies in  $\rm8-25\,M_\odot$ \citep{Arnett73, Miyaji80, Heger03}.
If we consider the initial mass function (IMF) by Salpeter,  described as
$\rm dN/dm=\xi_0m^{-2.35}$ \citep{Salpeter55}, where m is star mass in $\rm M_\odot$ units and $\rm \xi_0$ is normalization coefficient,
to calculate the boundary mass value for the high and low stellar masses, corresponding to the SNR-PSRs and non SNR-PSRs,
we obtain this critical mass to be  be at $\rm\sim12\,M_\odot$, as shown in Figure 6.
So, the SNRs  by  the low-mass ($\rm8-12\,M_\odot$) progenitor stars may be survived with  a shorter time around $\rm10-20\,Kyr$ \citep{Braun89}, which may be a reason for the  lots of young pulsars without SNRs.

\begin{figure}
\centering
\includegraphics[width=7.8cm]{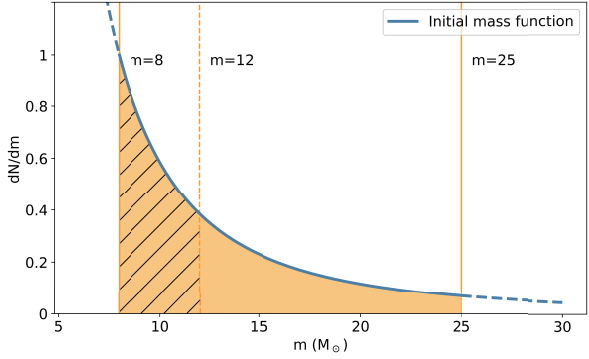}
\caption{The mass distribution of stars for SNR-PSRs and non SNR-PSRs  according to initial mass function (IMF) \citep{Salpeter55}.
The blue solid line is the curve of IMF.
The left and right orange solid lines are lower ($\rm8\,M_{\odot}$) and upper ($\rm25\,M_{\odot}$) mass limit for NS production, respectively.
The area  of  $\rm8-12\,M_{\odot}$  ($\rm12-25\,M_{\odot}$) stands for the progenitor stars for non-SNR-PSRs (SNR-PSRs).
 }
\label{fig6}
\end{figure}

The physical process of these two types of SNe can be described as the iron core collapse and electron capture, and the latter  is driven by the neutrino \citep{Janka12, Janka17}.
The iron core collapse is the dominant process of high-energy SNe that is generated from high-mass main sequence stars \citep{Heger03}.
The electron capture may be an explanation for the  low-energy SNe,  while the  degenerate oxygen-neon core is  collapsed to form the
NS\citep{Barkat74, Nomoto84}, and the mass range of these progenitor stars is about  $\rm\sim 8-10\,M_{\odot}$ \citep{Nomoto17, Leung20}.

Moreover, the mass boundary values given by  some researchers are similar to  ours, as $\rm\sim 12\,M_{\odot}$ \citep{Sugimoto80, Miyaji80}.
However  \cite{Nomoto84} and \cite{Heger03}  obtained  the boundary mass as  $\rm\sim 10\,M_{\odot}$.
It is remarked that our result is based on a statistics, but not a numerical calculation result from a stellar theoretical model.
In addition, because of less pulsar samples  at the young age ($<$10\,Kyr), the ratio in Eq.(3) may be biased, which will directly affect the mass boundary.

\hspace*{\fill} \\
Finally, the main conclusions are summarized below:
The 52 SNR-PSRs and 630 non SNR-PSRs  have been tested (K-S and M-W-W) and analyzed (cumulative number ratio), 
implying different $\dot P $ and other properties for the two sets, perhaps associated with the different mass ranges of their progenitor masses for  SN explosions.
The critical mass of different progenitor stars is estimated by the Salpeter initial mass function, obtained as $\rm 12\,M_{\odot}$.
The low-mass stars (high-mass) with $\rm\sim 8-12\,M_{\odot}$ ($\rm\sim 12-25\,M_{\odot}$) will generate the  low-energy (high-energy) SNe in the shorter (longer) SNR duration of about $\rm<10-20\,Kyr$ ($\rm >10-20\,Kyr$).
These conjectures can explain why many young radio pulsars are not seen inside SNRs.
In the future, with the   observations by  FAST \citep{Li18} and launch of James Webb Space Telescope (JWST) \citep{Gardner06}, more fainter and weaker pulsars and SNRs would be  discovered, which  will present the  better constraints on our conclusions.

\section*{Acknowledgments}

This work is supported by the National Natural Science Foundation of China (Grant No. 11988101, No. U1938117, No. U1731238, No. 11703003
and No. 11725313), the International Partnership Program of Chinese Academy of Sciences grant No. 114A11KYSB20160008, the National Key R\&D Program of China No. 2016YFA0400702, and the Guizhou Provincial Science and Technology Foundation (Grant No. [2020]1Y019).

\section*{Data Availability}
The data underlying this article are available in the references below:
(1) pulsars data are taken from ATNF Pulsar Catalogue, available at
https://www.atnf.csiro.au/research/pulsar/psrcat/;
(2) SNR data are taken from SNRcat, available at
http://snrcat.physics.umanitoba.ca/SNRtable.php.

\bsp

\clearpage
\section*{Appendix A: Derivation of period evolution equation}
Energy loss rate of a spin-powered pulsar can be expressed as
\begin{equation}
\begin{split}
\dot E = \frac{dE}{dt}=\frac{d(I\Omega^2/2)}{dt}=-I\Omega\dot\Omega,
\end{split}
\end{equation}
where $E$ is total energy of pulsar, $\dot E$ is energy loss rate, $\Omega$ is spin angular velocity, $\dot \Omega$ is rate of spin angular velocity, and $I$ is moment of inertia.
When the pulsar kinetic energy and the radiation energy loss rate are equal, then there is
\begin{equation}
\begin{split}
I\Omega\dot\Omega=K\Omega^{\rm n+1},
\end{split}
\end{equation}
where $K$ is coefficient,
We combined with $\Omega=2\pi/P$ and integrate both sides of above equation,
\begin{equation}
\begin{split}
\int_{P_0}^{P(t)}P^{\rm n-2}d\dot P=\int_{0}^{t}-\frac K I (2\pi)^{\rm n-1}dt.
\end{split}
\end{equation}
With assuming that the rate has been constant since pulsar birth, then spin period evolution equation can be written as
\begin{equation}
\begin{split}
P(t)=\left[({\rm n-1})P_0^{\rm n-2}\dot P_0 t +P_0^{\rm n-1} \right]^{1/(\rm n-1)}.
\end{split}
\end{equation}
In the above equation, $P_0$ and $\dot P_0$ is spin period and rate at present, respectively.
The spindown age of pulsar is $\tau=P /[({\rm n-1})\dot P]$,
which is different with characteristic age $\tau_c = P /(\rm2\dot P)$.
Then Eq.(7) can be rewritten as
\begin{equation}
\begin{split}
P(t)=P_0\left( \frac t \tau_0 + 1\right)^{1/(\rm n-1)}.
\end{split}
\end{equation}
When $t\gg\tau_0$, the equation can be simplified to
\begin{equation}
\begin{split}
P(t)\approx P_0 \left(\frac t \tau_0\right)^{1/(\rm n-1)}.
\end{split}
\end{equation}

\section*{Appendix B: Further tests of derivative of spin period, B-field and energy loss rate}

If the origin of pulsars that with or without SNRs are indeed different, then the distribution of other physical parameters should also be different.
Here we discuss the derivative of spin period ($\dot P$), B-field strength ($ B $ with n=3) and energy loss rate ($\dot E$) by applying K-S and M-W-W tests, and the  results are shown in Table 2, and the CDFs are shown in Figure 7, 8 and 9, where the distributions for two samples of SNR-PSRs and non SNR-PSRs are different respect to  these three parameters.
For $\dot P$, the two groups with the different ages  share the  significantly different distributions, which implies that the braking mechanisms of them perhaps are different.
The physical parameter distributions of two samples are quite different, which possibly could be ascribed to the different  origins of radio pulsars.

\begin{figure}
\centering
\includegraphics[width=7.8cm]{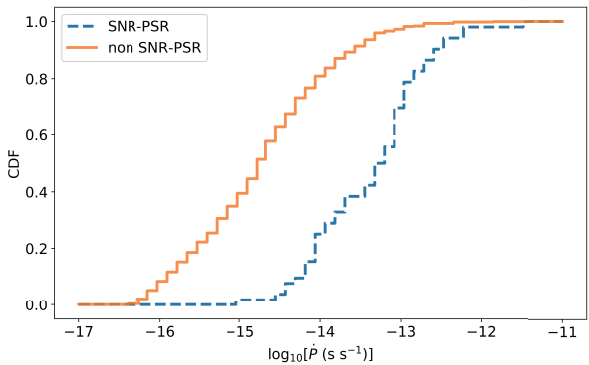}
\caption{The Cumulative distribution function (CDF) of derivative of spin period ($\dot P$) of SNR-PSRs and non SNR-PSRs. The  dashed line is for SNR-PSRs, and the  solid line is for non SNR-PSRs.}
\label{fig7}
\end{figure}
\begin{figure}
\centering
\includegraphics[width=7.8cm]{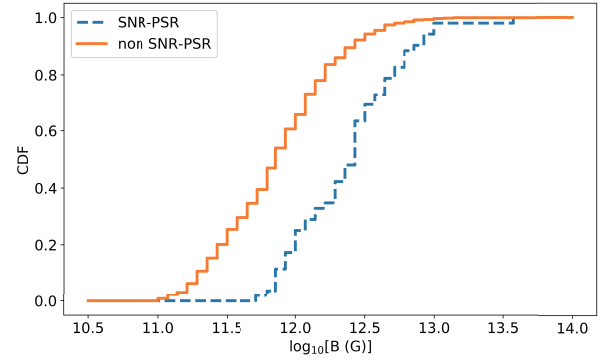}
\caption{The Cumulative distribution function (CDF) of surface magnetic field strength ($B$ with n=3) of SNR-PSRs and non SNR-PSRs. The dashed line is for SNR-PSRs, and the solid line is for non SNR-PSRs.}
\label{fig8}
\end{figure}
\begin{figure}
\centering
\includegraphics[width=7.8cm]{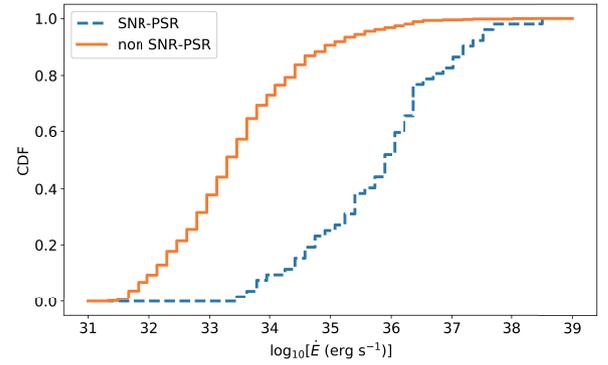}
\caption{The Cumulative distribution function (CDF) of spin down energy loss rate ($\dot E$) of SNR-PSRs and non SNR-PSRs. The  dashed line is for SNR-PSRs, and the   solid line is for non SNR-PSRs.}
\label{fig9}
\end{figure}

\begin{table}
\centering
\caption{P-values of K-S and M-W-W test for  different  parameters}
{
\begin{tabular}{@{}lcc@{}}
\hline
\noalign{\smallskip}
\bf Physical parameters$^a$ &\bf K-S test  &\bf M-W-W test \\
\hline
\noalign{\smallskip}
$P$ & $1.98\times10^{-12}$ & $5.85\times10^{-13}$ \\
$\dot P$ & $1.55\times10^{-15}$ & $6.41\times10^{-21}$ \\
$B$ & $3.24\times10^{-11}$ & $2.12\times10^{-14}$ \\
$\dot E$ & $1.55\times10^{-15}$ & $4.33\times10^{-24}$ \\
\hline
\end{tabular}
}
\label{tab2}
\begin{flushleft}
$^a$ $P$ is spin period, $\dot P$ is derivative of spin period, $B$ is surface magnetic field strength, and $\dot E$ is  energy loss rate  of radio pulsars.
\end{flushleft}
\end{table}

\section*{Appendix C: Tests of spin period and its derivative with the different ages}

\begin{figure*}
\centering
\includegraphics[width=15cm]{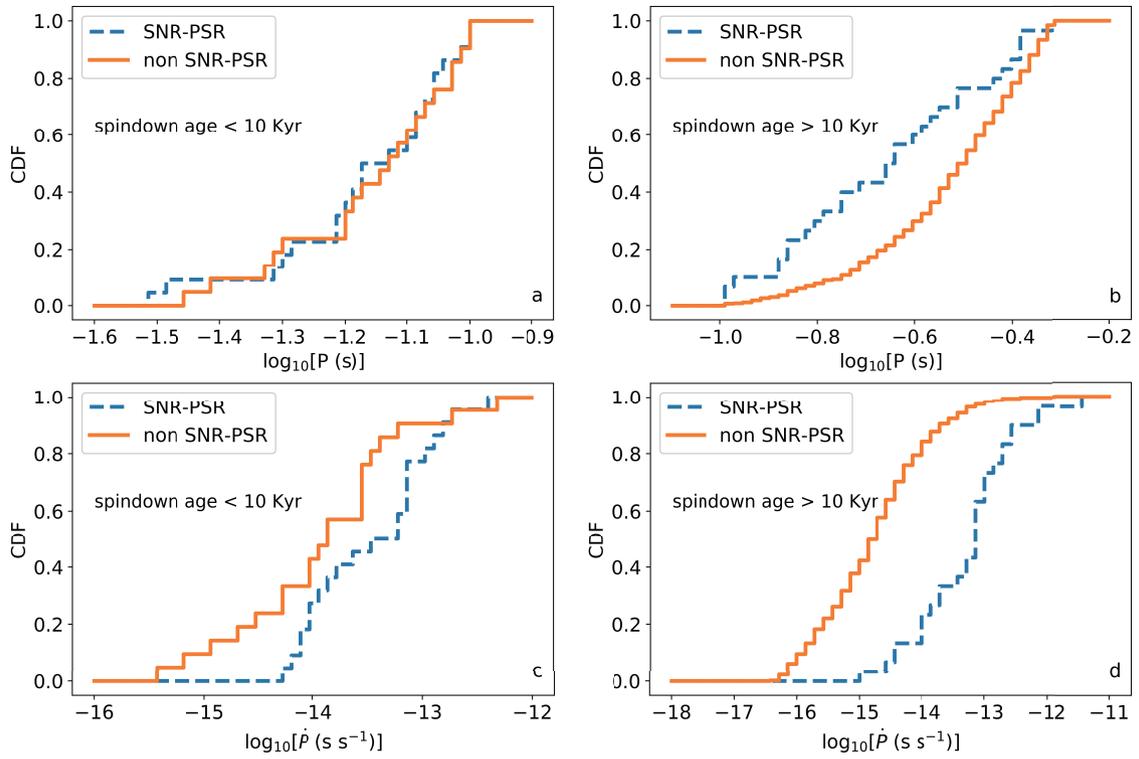}
\caption{The Cumulative distribution function (CDF) of spin period ($P$) and its derivative  ($\dot P$) of SNR-PSRs and non SNR-PSRs with the  spindown ages of  less and over than 10 Kyr.
The sub-figures of a and b are CDF of $P$, and the sub-figures of c and d are CDF of $\dot P$.
For all sub-figures, the  dashed line represents SNR-PSRs, and the  solid line is for non SNR-PSRs.
The text in each sub-figure shows their age ranges.
}
\label{fig10}
\end{figure*}

\begin{table}
\centering
\caption{P-values of K-S and M-W-W test for $P$ and $\dot P$ with different ages}
{
\begin{tabular}{@{}lccc@{}}
\hline
\noalign{\smallskip}
\bf Physical parameters$^a$ &\bf Spindown age &\bf K-S test  &\bf M-W-W test \\
\hline
\noalign{\smallskip}
\multirow{2}{*}{$P$} & <10 Kyr &0.99 & 0.85 \\
                     & >10 Kyr &$2.75\times10^{-3}$ & $4.83\times10^{-4}$ \\
 &&&\\
\multirow{2}{*}{$\dot P$} & <10 Kyr &$3.41\times10^{-2}$ & $5.34\times10^{-2}$$^b$ \\
                          & >10 Kyr & $4.38\times10^{-13}$ & $1.64\times10^{-13}$ \\
\hline
\end{tabular}
}
\label{tab3}
\begin{flushleft}
$^a$ $P$ is spin period, $\dot P$ is derivative of spin period.

$^b$ Although this value is slightly larger than 0.05, it may also imply that the two samples have different statistical distributions at a 90\% probability (e.g. if p-value is less than 0.1, it indicates that this test rejects the null hypothesis that the two samples have the same distribution at 10\% significance level).
\end{flushleft}
\end{table}

In this part, we plot distributions of $P$ with different age ranges of less and over than 10 Kyr in Figure 10 (subplots a \& b).
After K-S and M-W-W tests (Table 3), we find that when the spindown age is less than 10 Kyr the $P$ distributons of SNR-PSRs and non SNR-PSRs are  the same. 
But the distributions  of $P$ after 10 Kyr are different.
Meanwhile, distributions of $\dot P$ are also tested with the same age ranges as that of $P$ (less and more  than 10 Kyr) in Figure 10 (subplots c \& d).
Interestingly, regardless of the age ranges, the $\dot P$ distributions are different under K-S tests in Table 3.
The possible physical explanation is that although the initial $P$ distribution is the same, the initial $\dot P$ is different, which makes pulsars no longer have the same $P$ distribution after  evolution over 10 Kyr.
Thus, the above evidence supports that the differences between the SNR-PSRs and non SNR-PSRs are possibly the result of a combined effect of different mechanisms and evolution, but not just caused by age difference.

\label{lastpage}
\end{document}